\documentclass[final,5p,times,twocolumn]{elsarticle}

\usepackage{amssymb}
\usepackage{url}
\usepackage{amsmath}
\usepackage{multirow}
\usepackage{threeparttable}
\usepackage[colorlinks,
            linkcolor=blue,
            anchorcolor=blue,
            citecolor=green,
            urlcolor=red]{hyperref}

\journal{Information Sciences}

\begin{document}

\begin{frontmatter}



\title{Dual-Decoder Consistency via Pseudo-Labels Guided Data Augmentation for Semi-Supervised Medical Image Segmentation}


\author[label1,label2]{Yuanbin Chen}
\author[label1,label2]{Tao Wang}
\author[label1,label2]{Hui Tang}
\author[label1,label2]{Longxuan Zhao}
\author[label1,label2]{Ruige Zong}
\author[label3,label4]{Shun Chen}
\author[label5]{Tao Tan}
\author[label1,label2]{Xinlin Zhang\corref{cor}}
\ead{xinlin1219@gmail.com}
\author[label1,label2]{Tong Tong\corref{cor}}
\ead{ttraveltong@gmail.com}

\cortext[cor]{Corresponding author}

\affiliation[label1]{organization={College of Physics and                    Information Engineering},
            addressline={Fuzhou University},
            city={Fuzhou},
            state={China}}

\affiliation[label2]{organization={Fujian Key Lab of Medical                 Instrumentation \& Pharmaceutical Technology},
            addressline={Fuzhou University},
            city={Fuzhou},
            state={China}}

\affiliation[label3]{organization={Department of Ultrasound},
            addressline={Fujian Medical University Union Hospital},
            city={Fuzhou},
            state={China}}

\affiliation[label4]{organization={Fujian Medical Ultrasound Research Institute},
            city={Fuzhou},
            state={China}}

\affiliation[label5]{organization={Faculty of Applied Science},
            addressline={Macao Polytechnic University},
            city={Macao},
            state={China}}

\begin{abstract}
While supervised learning has achieved remarkable success, obtaining large-scale labeled datasets in biomedical imaging is often impractical due to high costs and the time-consuming annotations required from radiologists. Semi-supervised learning emerges as an effective strategy to overcome this limitation by leveraging useful information from unlabeled datasets. In this paper, we present a novel semi-supervised learning method, Dual-Decoder Consistency via Pseudo-Labels Guided Data Augmentation (DCPA), for medical image segmentation. We devise a consistency regularization to promote consistent representations during the training process. Specifically, we use distinct decoders for student and teacher networks while maintain the same encoder. Moreover, to learn from unlabeled data, we create pseudo-labels generated by the teacher networks and augment the training data with the pseudo-labels. Both techniques contribute to enhancing the performance of the proposed method. The method is evaluated on three representative medical image segmentation datasets. Comprehensive comparisons with state-of-the-art semi-supervised medical image segmentation methods were conducted under typical scenarios, utilizing $10$\% and $20$\% labeled data, as well as in the extreme scenario of only $5$\% labeled data. The experimental results consistently demonstrate the superior performance of our method compared to other methods across the three semi-supervised settings. The source code is publicly available at \url{https://github.com/BinYCn/DCPA.git}.
\end{abstract}




\end{frontmatter}


\section{Introduction}
\label{sec1}
Image segmentation plays a crucial and indispensable role in medical image analysis by annotating structures of interest in various imaging modalities, such as computerized tomography (CT) or magnetic resonance imaging (MRI). This annotation is a valuable asset for the continuous monitoring and accurate diagnosis of diseases. It is remarkable to see the significant progress that has been made in leveraging deep learning for medical image segmentation.
 Among numerous deep learning methods, convolutional neural network (CNN) techniques have demonstrated cutting-edge performance and widespread adoption \cite{tang2024htc}. For instance, fully convolutional networks \cite{long2015fully} and encoder-decoder networks, such as U-Net \cite{ronneberger2015u} and V-Net \cite{milletari2016v}, have successively been proposed for this task. The effectiveness of these methodologies is intricately tied to the availability of high-quality annotated datasets. However, acquiring large-scale labeled datasets in medical imaging is often impractical due to the associated high costs and the time-intensive nature of annotations performed exclusively by experienced radiologists.

Semi-supervised learning emerges as an effective approach to address the demand for numerous labeled data in supervised learning while maintaining promising segmentation accuracy. 
These methods extract valuable information from a vast amount of unlabeled data, thereby enhancing segmentation accuracy and establishing themselves as prominent techniques in medical image segmentation \cite{you2022simcvd}. A critical research problems in semi-supervised learning is developing effective strategies for supervising unlabeled data, which already gives rise to numerous methods that adeptly exploit the latent potential of unlabeled data. Semi-supervised methods can be primarily classified into three distinct categories: (1) Consistency regularization methods \cite{laine2017temporal,tarvainen2017mean,sajjadi2016regularization}, which induce the model to exhibit consistent output distributions when subjected to input perturbations. By subjecting the unlabeled data to diverse perturbations or employing augmentation techniques, the model is compelled to generate coherent predictions, thereby augmenting its generalization capability. (2) Entropy minimization methods \cite{abdar2021review,berthelot2019mixmatch,chen2020mixtext}, which are grounded in the clustering assumption, actively encourage the model to make confident predictions on unlabeled data. Through the minimization of entropy in the predicted probability distribution, the model acquires an increased degree of certainty, consequently fortifying its learning effectiveness. (3) Pseudo-labeling methods \cite{cascantebonilla2020curriculum,lee2013pseudo,rizve2021defense}, serving as a mechanism to enrich the training set, enhance model performance by generating pseudo-labels from the unlabeled data.

Semi-supervised learning methods have achieved promising performance in medical image segmentation, but they also face some obstacles. \cite{zhao2023rcps}. Firstly, given varying advantages and limitations of different methods, it is a crucial task in selecting an appropriate method and/or an effective regularization. Besides, the large amount of unlabeled data used in the learning process may cause a potential imbalance problem. This problem can easily lead to the model being biased towards unlabeled data, thereby reducing the performance of semi-supervised learning methods. Third, the intricate and diverse characteristics present in medical images, such as intensity variations among imaging modalities, increase the difficulty of capturing underlying patterns and achieving generalization across different datasets. In light of these limitations, the development of robust semi-supervised learning methods remains a challenging yet crucial research problem.

This paper addresses the issues of medical image segmentation through the introduction of a semi-supervised learning technique known as Dual-Decoder Consistency via Pseudo-Labels Guided Data Augmentation (DCPA). The proposed approach is designed to comprehensively extract information from both labeled and unlabeled data, aiming to enhance the efficacy of unlabeled regions through two key mechanisms. The first mechanism involves Pseudo-Labels generation and guided data augmentation. Within the DCPA framework, the teacher model is responsible for generating pseudo-labels for the unlabeled dataset. Following this generation, a sophisticated data augmentation process is employed, merging unlabeled data with its labeled counterpart using the Mixup technique \cite{zhang2018mixup}. This process not only enriches the dataset but also ensures that the generated pseudo-labels serve as definitive labels for the mixed dataset. The second mechanism is consistency via Dual-Decoders. The DCPA architecture comprises both teacher and student models, each equipped with a shared encoder and dual-decoders, each employing distinct upsample strategies. To ensure the reliability and consistency of the model's predictions, the outputs of the student model's dual-decoders are harmonized through meticulously designed consistency constraints.

The main contributions of this paper are as follows:
\begin{itemize}
	\item We introduce a novel semi-supervised medical image segmentation paradigm, integrating pseudo-labels and consistency. This innovative approach adeptly harnesses the latent potential of unlabeled data, resulting in significant enhancements in the efficacy of semi-supervised image segmentation.

	\item We propose a novel dual-decoder network, underpinned by the mean-teacher model. The integration of a consistency scheme guides the model's training trajectory, facilitating the extraction of holistic and universal feature representations.

	\item Comprehensive evaluations of the proposed method were conducted on three challenging public medical image segmentation benchmarks. The experimental results showcase the superiority of the proposed method, as evidenced by its outperformance of state-of-the-art methods.
\end{itemize}

\section{Related work}

\subsection{Supervised medical image segmentation}

In recent years, the superior learning capability of deep learning has inspired a fruitful of supervised learning based methods for medical image segmentation\cite{chen2023cotrfuse,wang2022net}. The application of encoder-decoder based architectures (e.g., UNet \cite{ronneberger2015u} and nnUNet \cite{isensee2021nnu}) has significantly improved the accuracy of various modality and segmentation tasks, and there have been many researches on obtaining better performance by improving the network architecture. M$^{2}$SNet \cite{zhao2023m2snet} proposes a simple and generalized multi-scale subtraction network. This network efficiently acquires multiscale complementary information across different levels, progressing from lower to higher orders, comprehensively enhances the perception of organs or lesion regions. Furthermore, the Transformer model \cite{vaswani2017attention} has attracted much attention in the field of medical image segmentation due to its excellent ability in modelling global context. DAEFormer \cite{azad2023dae} designs a novel Transformer architecture based on a dual attention mechanism guided to capture spatial and channel relationships across the entire feature dimensions while maintaining the computational efficiency. Additionally, the method redesigns jump connection paths by incorporating cross-attention modules, augmenting the model's localization performance. Some researchers have innovatively combined CNN with Transformers to capitalize on the strengths of both paradigms. DA-TransUNet \cite{sun2023transunet} is an illustrative example, leveraging the attention mechanism of Transformer and the multifaceted feature extraction of the Dual Attention Block. This amalgamation efficiently combines global, local, and multi-scale features to enhance medical image segmentation. Supervised learning based medical image segmentation methods obtain impressive performance, but requires large amount of manually labelled data for training.

\begin{figure*}[!t]
\centerline{\includegraphics[width=\textwidth]{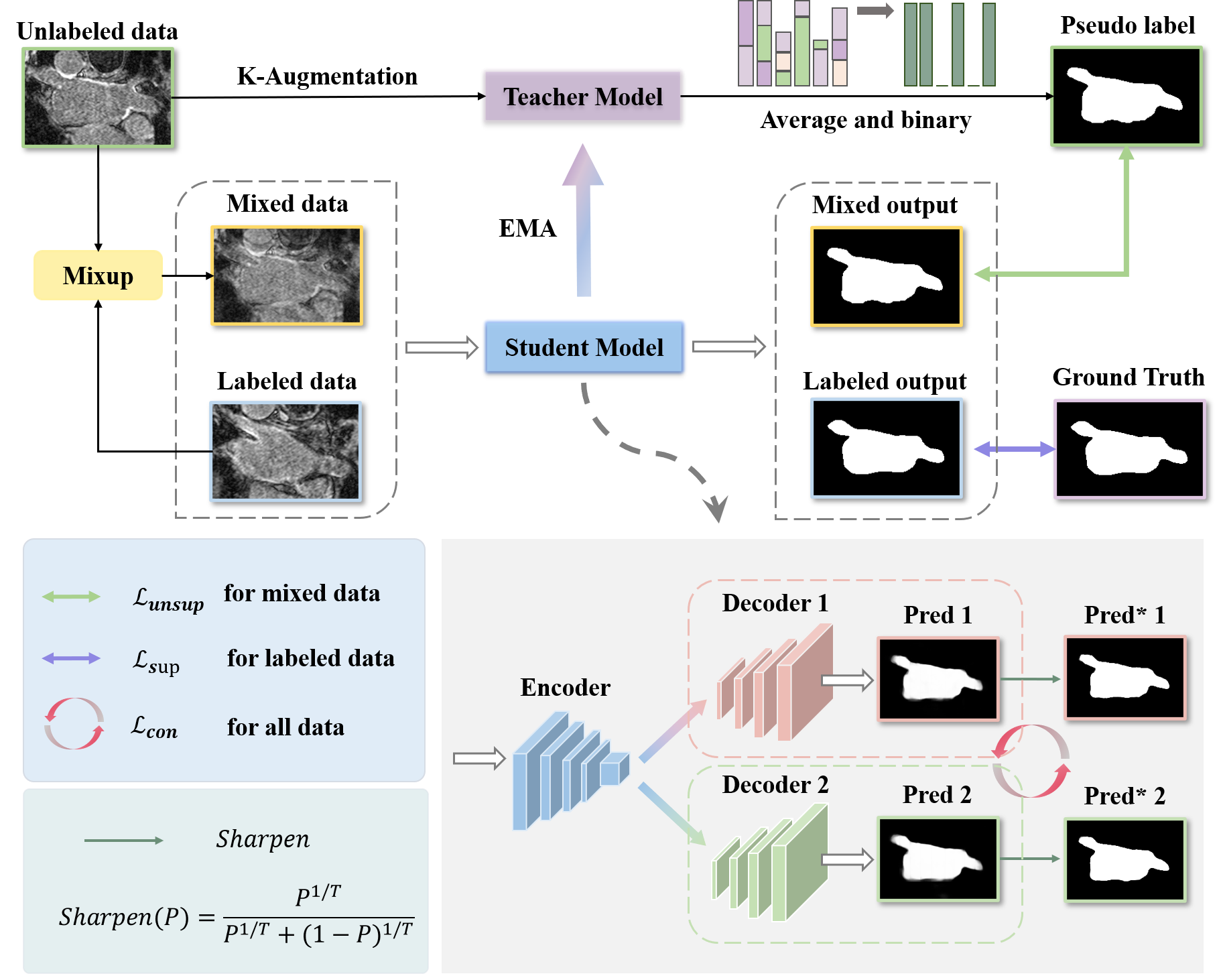}}
\caption{Overview of the proposed DCPA method. Detailed descriptions of both the teacher model and the student model are provided beneath the figure. Notably, Decoder 1 and Decoder 2 correspond to two distinct decoders employing different upsampling strategies. The circular annotations of $L_{con}$ indicate the computation of consistency loss between Pred 1 and Pred* 2, as well as between Pred 2 and Pred* 1. EMA stands for exponential moving average.}
\label{DCPA}
\end{figure*}

\subsection{Semi-supervised medical image segmentation}
Semi-supervised learning has gained prominence as an effective approach to address the challenges associated with acquiring large-scale labeled data, particularly in the field of medical image segmentation. To optimally leverage information from unlabeled data, researchers have proposed numerous techniques grounded in diverse principles. A prevalent strategy involves the use of consistency regularization, where the model is trained to maintain consistency in its predictions across varied tasks or representations \cite{li2020shape,luo2021semi,luo2022semi}. This consistency is emphasized even when minor perturbations, such as noise or data augmentations, are introduced to the input \cite{li2020transformation,tarvainen2017mean,yu2019uncertainty}. For instance, TCSM-V2 \cite{li2020transformation} incorporated multiple data and model perturbations—such as flipping, rotation, resizing, noise addition, and dropout—within the mean-teacher framework \cite{tarvainen2017mean} to ensure consistent model predictions across different input variations. Additionally, Li et al. \cite{li2020shape} utilized a multi-task network structure tailored for image segmentation and signed distance map regression, integrating shape and position priors. They incorporated a discriminator as a regularization component, enhancing the stability and robustness of segmentation outcomes. Another widely adopted approach is entropy minimization, prompting the model to produce confident predictions for unlabeled data by reducing the entropy of the model's output probability distribution \cite{hang2020local,wu2022mutual}. For example, Hang et al. \cite{hang2020local} integrated entropy minimization into the student network, enabling it to generate confident predictions for unlabeled images. Another category of methods is based on pseudo label techniques. These methods generate pseudo-labels for unlabeled data, guiding the model's learning trajectory. Wang et al. \cite{wang2021neighbor} proposed the integration of a trust module to reassess pseudo-labels derived from model outputs, setting a threshold to cherry-pick high-confidence predictions. Meanwhile, Li et al. \cite{li2021self} unveiled a self-ensembling strategy that employs Exponential Moving Average (EMA) to yield precise and dependable predictions, addressing challenges linked with noisy and fluctuating pseudo-labels.

The previously mentioned methods have certain drawbacks. Some necessitate intricate network structures and additional training components, thereby increasing computational and resource demands. Others exhibit suboptimal performance when dealing with noisy or uncertain pseudo-labeling, resulting in instability or a decline in performance. In response to these challenges and limitations, we have introduced a new method building upon previous work. This method integrates pseudo-labels, consistency regularization, and data augmentation within a model based on the mean-teacher framework for semi-supervised medical image segmentation, offering a solution that is both straightforward and highly effective.

\section{Method}
To provide a clear and precise representation of the proposed method, we introduce the following notations. We partition the training dataset into two subsets: a labeled set, $D^N_L$, with $N$ samples, and an unlabeled set, $D^M_U$, with $M$ samples. For an image $x_i \in D^N_L$, its corresponding ground truth label $y_i$ is known. However, for an image $x_i \in D^M_U$, the ground truth label is absent. Thus, the datasets are defined as:
\begin{equation}\nonumber
	\begin{aligned}
		D^N_L &= \{x^i_l, y^i_l | i = 1, ..., N\}, \\
		D^M_U &= \{x^i_u | i = 1, ..., M\}.
	\end{aligned}
\end{equation}
The probability map generated for $x$ is represented by $p(y_{pred} | x ; \theta)$, where $\theta$ signifies the parameters of the segmentation model $f(\theta)$.

\subsection{Model architecture}
In this paper, we introduce a novel semi-supervised medical image segmentation method, referred to as DCPA, as illustrated in Fig. \ref{DCPA}. The DCPA framework is built upon two core components which can fully use both labeled and unlabeled data. The first component focuses on data augmentation. We harness the potential of unlabeled data by generating pseudo-labels. To further enhance the training dataset, we employ Mixup, an effective augmentation technique, which merges these pseudo-labeled data with the original labeled samples. This augmented dataset then serves as the foundation for the subsequent training phase. The second component revolves around the architecture and training of the student and teacher models. Both models share a consistent structure, featuring a unified encoder and a dual-decoder, each equipped with distinct up-sampling strategies. However, they differ in the training strategies. The student model is trained using our proposed loss function, while the teacher model, in contrast, adopts the EMA approach, adjusting the weights based on the student model's performance.

Subsequent sections detailly introduce the two components of our methodology.

\subsection{Pseudo-Labels Guided Data Augmentation}
\subsubsection{Pseudo-Labels Generation}
The weak data augmentation techniques, such as random perturbations, are firstly applied to our dataset. This augmentation is applied differently for labeled and unlabeled images. For every batch of labeled images, denoted as \(X^b_L \in D^N_L\), we generate a single augmented version, represented as $\hat{X}^b_L = Augment(X^b_U)$. In contrast, for each batch of unlabeled images $X^b_U \in D^M_U$, we produce $K$ augmented versions, given by $\hat{X}^{b ,k}_U= Augment(X^b_U)$ for $k \in \{1,...,K\}$.

Once the data is augmented, we employ the teacher model to compute the average predicted segmentation distribution, $\overline{P}^b$, from the augmented unlabeled samples. This average distribution is then subjected to a binary operation to generate the pseudo-labels, $P^b$. The binary operation is conducted based on a threshold, $H$, which is commonly set to $0.5$. The specific computational formulas are:
\begin{equation}
\overline{P}^b = \frac{1}{K} \sum^K_{k=1} p(y_{pred} | \hat{X}^{b,k}_U ; \theta_t),
\end{equation}
\begin{equation}
P^b(x, y) = \begin{cases}
            1, & \text{if } \overline{P}^b(x, y) \geq H, \\
            0, & \text{otherwise.}
            \end{cases}
\end{equation}
where, $\theta_t$ represents the parameters of the teacher model, and $P^b(x, y)$ denotes the pixel value at the coordinate $(x, y)$ within the pseudo-labels. If the pixel value in $\overline{P}^b$ meets or exceeds the threshold $H$, its corresponding value in $P^b$ is assigned as 1; otherwise, it remains 0. These pseudo-labels, $P^b$, are subsequently integrated into the unsupervised loss term.

\begin{figure}[t]
\centerline{\includegraphics[width=\columnwidth]{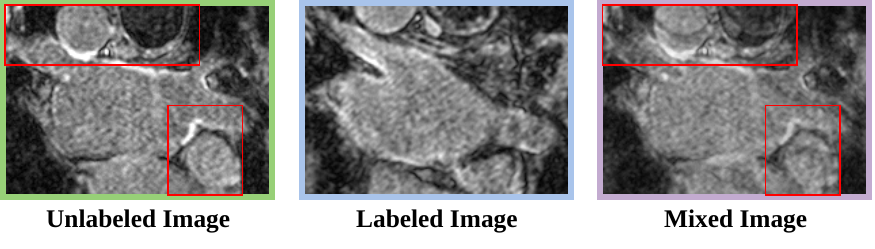}}
\caption{Illustration of the effects of Mixup. The red rectangles within the image highlight areas where noticeable alterations are observed before and after applying Mixup.}
\label{mix}
\end{figure}

\subsubsection{Data Augmentation with Mixup}
Drawing inspiration from Mixmatch \cite{berthelot2019mixmatch}, we incorporate Mixup \cite{zhang2018mixup} to combine augmented unlabeled images with augmented labeled images. This fusion enhances data diversity through augmentation. Given a pair of images, $x_1$ and $x_2$, the resultant mixed image, $x'$, is formulated as:
\begin{equation}
\lambda \sim Random(Beta(\alpha,\alpha)),
\label{eq3}
\end{equation}
\begin{equation}
\lambda' = Max(\lambda,1-\lambda),
\label{eq4}
\end{equation}
\begin{equation}
x' = \lambda'x_1 + (1-\lambda')x_2,
\label{eq5}
\end{equation}
where $\alpha$ is a hyperparameter, and $Beta$ denotes the Beta distribution. Note that we do not apply Mixup \cite{zhang2018mixup} to labeled images due to the inherent complexities of medical images, which often exhibit significant variations in morphology, size, and anatomical positioning.

Fig. \ref{mix} shows the results of applying Mixup \cite{zhang2018mixup} to a pair of images. Post-Mixup, the resultant image predominantly retains the semantic essence of the unlabeled data. However, the inclusion of labeled data introduces subtle modifications, especially in areas highlighted by the red rectangle. Despite these alterations, the core semantic information remains largely intact. Thus, we designate the pseudo-labels $P^b$ as the label for the mixed image.

To streamline the Mixup process, we initiate by shuffling the augmented labeled images, $\hat{X}^b_L$, to form set $W^b$. Subsequently, Mixup is applied to merge the augmented unlabeled images, $\hat{X}^{b ,k}_U$, with set $W^b$, culminating in the creation of mixed images $M^{b ,k}$. These mixed images are then paired with pseudo-labels $P^b$ to produce the mixed data $D_M$. The corresponding formulas are:
\begin{equation}
W^b = Shuffle(\hat{X}^b_L),
\label{eq6}
\end{equation}
\begin{equation}
M^{b ,k} = \left( Mixup(\hat{X}^{b ,k}_U, W^b_{i\%|\hat{X}^b_L|}) ; i \in (1,...,|\hat{X}^{b ,k}_U|) \right),
\label{eq7}
\end{equation}
\begin{equation}
D_M = \left( (M^{b ,k}, P^b) ; b \in (1,...,B), k \in (1,...,K) \right),
\label{eq8}
\end{equation}
where the symbol $\%$ signifies the modulo operation.

\subsection{Dual-Decoder Consistency}
\subsubsection{Dual-Decoder architecture based on the mean-teacher framework}
Fig. \ref{DCPA} illustrates the architecture of both the student and teacher models, which are constructed upon the V-Net backbone. This design comprises a shared encoder and a dual-decoder, each with distinct strategies: the transposed convolutional layer and the linear interpolation layer. The purpose of this dual-decoder approach is to exploit the differences between the decoder outputs, thereby capturing and representing the model's inherent uncertainty.

To enhance the quality of pseudo-labels derived from unlabeled data, we adopt the Exponential Moving Average (EMA) approach. This method updates the teacher model's parameters based on the student model's parameters, integrating historical states throughout the learning trajectory. The effectiveness of this approach is supported by \cite{tarvainen2017mean}. The update rule for the teacher model's parameters at a given time step, represented as $\theta^{'}_t$, is:
\begin{equation}
\theta^{'}_t = \alpha \theta^{'}_{t-1} + (1-\alpha) \theta_t,
\label{eq9}
\end{equation}
where $\theta^{'}_{t-1}$ denotes the teacher model's historical parameters, while $\theta_t$ represents the student model's current parameters. The hyperparameter $\alpha$ serves as a smoothing coefficient, balancing the relationship between the teacher and student models.

\subsubsection{Sharpening function}
\begin{figure}[!t]
\centerline{\includegraphics[width=\columnwidth]{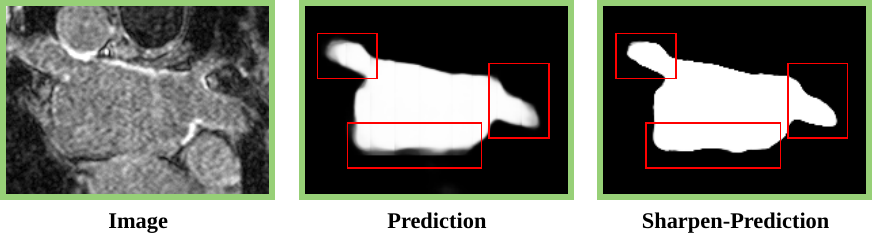}}
\caption{Demonstration of results after post-sharpening. The red rectangles spotlight regions with pronounced alterations pre and post-sharpening.}
\label{sharp}
\end{figure}

To maximize the utility of unlabeled data, we introduce a sharpening function. This function aims to reduce the entropy of the probability maps generated by the student model, resulting in soft-pseudo-labels $Pred^*_i$:
\begin{equation}
Pred^*_i = \frac{Pred_i^{\frac{1}{T}}}{Pred_i^{\frac{1}{T}} + \left(1-Pred_i\right)^{\frac{1}{T}}}, i \in (1,2),
\label{eq10}
\end{equation}

where the hyperparameter $T$ controls the sharpening temperature. As $T$ approaches zero, the output of $Pred^*_i$ converges to a Dirac ("one-hot") distribution, prompting the model to yield predictions with diminished entropy. The sharpening operation's impact is shown in Fig. \ref{sharp}, where segmentation boundaries are accentuated and image blurriness is reduced, especially in regions delineated by the red rectangle.

Building upon the sharpening process, we then employ the derived soft-pseudo-labels $Pred^*_1$ and $Pred^*_2$ to supervise the probability maps $Pred_2$ and $Pred_1$, respectively. This step enforces a consistency constraint, bridging the gap between decoder outputs and guiding the model's learning direction.

\subsection{Loss Function}
The overall loss function has three key components. The first component, $L_{con}$, is introduced to minimize the discrepancies between different decoder outputs. Here the final output of the student model is the average of the two probability maps, i.e., $Pred=\frac{1}{2} \cdot (Pred_1 + Pred_2)$. The second component, $L_{sup}$, minimizes the error of the prediction that has ground-truth labels while the third component, $L_{unsup}$, minimizes the error of the prediction that has pseudo-labels $P$. The overall loss is given by
\begin{equation}
	L_{total} = L_{sup} + L_{unsup} + \lambda L_{con},
	\label{eq11}
\end{equation}
where
\begin{equation}
\begin{aligned}
	& L_{unsup} = L_{dice}(Pred, P), \\
	& L_{sup} = L_{dice}(Pred, Y) + L_{ce}(Pred, Y), \\
	& L_{con} = L_{mse}(Pred_1, Pred^*_2) + L_{mse}(Pred_2, Pred^*_1),
\label{eq12}
\end{aligned}
\end{equation}
and $L_{mse}$, $L_{ce}$ and $L_{dice}$ denote mean squared error, cross-entropy and Dice \cite{milletari2016v} respectively. The parameter $\lambda$ is a widely used time-dependent Gaussian warming-up function \cite{laine2017temporal} given by $\lambda(t) = w_{max} \cdot \exp ( -5(1-\frac{t}{t_{max}})^2 )$, and $w_{max}$, $t$, and $t_{max}$ denotes the ultimate regularization weight, the current training step, and the maximum training step, respectively.

\section{Experiments}
In this section, three representative medical image datasets were utilized to extensively evaluate the performance and effectiveness of the proposed method.

\subsection{Dataset and Preprocessing}

\subsubsection{Pancreas-CT dataset} The pancreas-CT dataset \cite{clark2013cancer} utilized in this study was obtained from the National Institutes of Health Clinical Center. The dataset comprises $82$ 3D abdominal contrast-enhanced CT scans acquired using Philips and Siemens MDCT scanners. These scans have a fixed in-plane resolution of $512 \times 512$ and varying intra-slice spacing ranging from $1.5$ to $2.5$ mm. To ensure consistency, a voxel value clipping operation \cite{luo2021semi} was employed, restricting values to the range of $[-125, 275]$ Hounsfield Units (HU). Subsequently, the data underwent resampling to achieve an isotropic resolution of $1.0 \times 1.0 \times 1.0$ mm. Following a similar approach to \cite{luo2022semi, wu2022mutual}, we selected $62$ samples for the training set and $20$ samples for the testing set.

\subsubsection{LA dataset} The LA (Left Atrium) dataset \cite{xiong2021global}, established as a benchmark dataset for the $2018$ Atrial Segmentation Challenge, encompasses $154$ gadolinium-enhanced MR scans with an isotropic resolution of $0.625 \times 0.625 \times 0.625$ $mm^3$. Given that this dataset lacks labels for its test set, our study utilized only its labeled samples, amounting to a total of $100$ samples. Specifically, $80$ samples were allocated to the training set, while $20$ samples were reserved for the testing set.

\subsubsection{ACDC dataset} The ACDC (Automated Cardiac Diagnosis Challenge) dataset \cite{bernard2018deep} was curated from real clinical exams conducted at the University Hospital of Dijon. This dataset comprises cardiac MR imaging samples obtained from $100$ patients, featuring a short-axis in-plane spatial resolution ranging from $0.83$ to $1.75$ $mm^2/pixel$. Significantly, each patient scan in the dataset is meticulously annotated with mask information corresponding to the left ventricle (LV), right ventricle (RV), and myocardium (MYO). Following the approach in existing work \cite{luo2021semi}, we designated $70$ samples for the training set, $10$ samples for the validation set, and $20$ samples for the testing set.

\subsection{Implementation Details}

\subsubsection{3D Segmentation} The typical image preprocessing techniques \cite{luo2021semi,wu2022mutual,yu2019uncertainty}, including cropping, normalization, and unit variance, were applied to the images before feeding them into the neural networks. Due to the computational demands associated with 3D training, a patch-based approach was employed during the training phase, where the training data was divided into small patches. Specifically, a patch size of $112 \times 112 \times 80$ was used for the LA dataset, while a patch size of $96 \times 96 \times 96$ was used for the pancreas-CT dataset. Moreover, data augmentation techniques such as 2D rotation and flip operations were applied to augment the LA dataset.

For semi-supervised learning experiments, a subset of the training data was used as the labeled training set, while the rest served as the unlabeled training set. Specifically, we considered three different settings, i.e., training with $5$\%, $10$\%, and $20$\% labeled data, respectively. An unusual setting with only $5$\% labeled data was adopted to further evaluate the performance of the proposed method under the extreme scenario. For the proposed method, the V-Net was utilized as the network backbone. A batch size of $4$ was employed, including two labeled patches and two unlabeled patches. The 3D DCPA model underwent training for a total of $15,000$ iterations. During the inference stage, a patch-based pipeline with a sliding window strategy was employed to combine predictions from patches and generate the final segmentation outputs.

\begin{table*}[ht]\small
\caption{Comparison with the SOTA methods on the Pancreas-CT dataset.}
\label{tab1}
\begin{center}
\begin{tabular}{llclclccccccc}
\hline
\multicolumn{1}{c}{\multirow{2}{*}{\textbf{Method}}} &  & \multicolumn{3}{c}{\textbf{Scans used}}                                      &  & \multicolumn{7}{c}{\textbf{Metrics}}                                                                \\ \cline{3-5} \cline{7-13} 
\multicolumn{1}{c}{}                                 &  & Labeled                   & \multicolumn{1}{c}{} & Unlabeled                 &  & Dice(\%)$\uparrow$       &           & Jaccard(\%)$\uparrow$    &           & 95HD(voxel)$\downarrow$   &           & ASD(voxel)$\downarrow$    \\ \hline
V-Net                                                &  & 3(5\%)                    & \multicolumn{1}{c}{} & 0                         &  & 29.32          &           & 19.61          &           & 43.67         &           & 15.42         \\
V-Net                                                &  & 6(10\%)                   & \multicolumn{1}{c}{} & 0                         &  & 54.94          &           & 40.87          &           & 47.48         &           & 17.43         \\
V-Net                                                &  & 12(20\%)                  & \multicolumn{1}{c}{} & 0                         &  & 71.52          &           & 57.68          &           & 18.12         &           & 5.41          \\
V-Net                                                &  & 62(All)                   & \multicolumn{1}{c}{} & 0                         &  & 83.48          &           & 71.99          &           & 4.44          &           & 1.26          \\ \hline
UA-MT (MICCAI'19)                                               &  & \multirow{7}{*}{3(5\%)}   &                      & \multirow{7}{*}{59(95\%)} &  & 43.15          &           & 29.07          &           & 51.96         &           & 20.00         \\
SASSNet (MICCAI'20)                                             &  &                           &                      &                           &  & 41.48          &           & 27.98          &           & 47.48         &           & 18.36         \\
DTC (AAAI'21)                                                 &  &                           &                      &                           &  & 47.57          &           & 33.41          &           & 44.17         &           & 15.31         \\
URPC (MIA'22)                                                &  &                           &                      &                           &  & 45.94          &           & 34.14          &           & 48.80         &           & 23.03         \\
SS-Net (MICCAI'22)                                              &  &                           &                      &                           &  & 41.39          &           & 27.65          &           & 52.12         &           & 19.37         \\
MC-Net+ (MIA'22)                                             &  &                           &                      &                           &  & 32.45          &           & 21.22          &           & 58.57         &           & 24.84         \\
\textbf{Ours}                                        &  &                           &                      &                           &  & \textbf{79.16} &           & \textbf{65.86} &           & \textbf{5.88} &           & \textbf{1.69} \\ \hline
UA-MT (MICCAI'19)                                                &  & \multirow{7}{*}{6(10\%)}  &                      & \multirow{7}{*}{56(90\%)} &  & 66.44          & \textbf{} & 52.02          & \textbf{} & 17.04         & \textbf{} & 3.03          \\
SASSNet (MICCAI'20)                                             &  &                           &                      &                           &  & 68.97          & \textbf{} & 54.29          & \textbf{} & 18.83         & \textbf{} & 1.96          \\
DTC (AAAI'21)                                                 &  &                           &                      &                           &  & 66.58          &           & 51.79          &           & 15.46         &           & 4.16          \\
URPC (MIA'22)                                                &  &                           &                      &                           &  & 73.53          &           & 59.44          &           & 22.57         &           & 7.85          \\
SS-Net (MICCAI'22)                                              &  &                           &                      &                           &  & 73.44          &           & 58.82          &           & 12.56         &           & 2.91          \\
MC-Net+ (MIA'22)                                             &  &                           &                      &                           &  & 70.00          &           & 55.66          &           & 16.03         &           & 3.87          \\
\textbf{Ours}                                        &  &                           &                      &                           &  & \textbf{81.34} &           & \textbf{68.97} &           & \textbf{5.40} &           & \textbf{1.69} \\ \hline
UA-MT (MICCAI'19)                                               &  & \multirow{7}{*}{12(20\%)} &                      & \multirow{7}{*}{50(80\%)} &  & 76.10          &           & 62.62          &           & 10.84         &           & 2.43          \\
SASSNet (MICCAI'19)                                             &  &                           &                      &                           &  & 76.39          & \textbf{} & 63.17          & \textbf{} & 11.06         & \textbf{} & 1.42          \\
DTC (AAAI'21)                                                 &  &                           &                      &                           &  & 76.27          &           & 62.82          &           & 8.70          &           & 2.20          \\
URPC (MIA'22)                                                &  &                           &                      &                           &  & 80.02          &           & 67.30          &           & 8.51          &           & 1.98          \\
SS-Net (MICCAI'22)                                              &  &                           &                      &                           &  & 78.68          &           & 65.96          &           & 9.74          &           & 1.91          \\
MC-Net+ (MIA'22)                                             &  &                           &                      &                           &  & 79.37          &           & 66.83          &           & 8.52          &           & 2.03          \\
\textbf{Ours}                                        &  &                           &                      &                           &  & \textbf{82.89} &           & \textbf{71.17} &           & \textbf{4.54} &           & \textbf{1.23} \\ \hline
\end{tabular}
\end{center}
\end{table*}

\begin{figure*}[!t]
\centerline{\includegraphics[width=\textwidth]{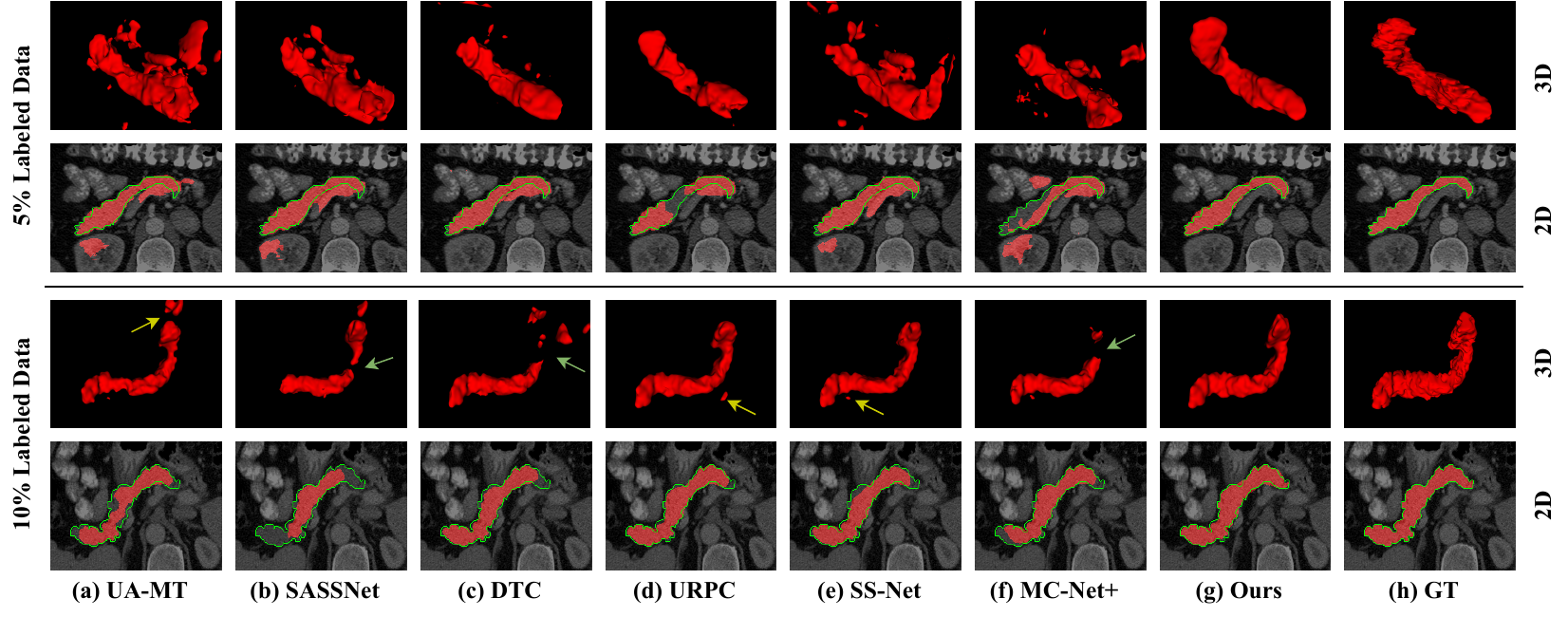}}
\caption{2D and 3D views of the segmentation results on the Pancreas-CT dataset. For better visualization, we delineate the segmentation outline of the ground truth (green) and overlay it over the predictions.}
\label{PA}
\end{figure*}

\subsubsection{2D Segmentation} On the ACDC dataset, normalization was performed to achieve zero mean and unit variance for the samples. Additionally, to enhance dataset diversity, data were augmented by randomly applying rotations and flips. Similar to the 3D segmentation experiments, a subset of training data was randomly selected as the labeled training set, while the rest served as the unlabeled training set. Specifically, three different settings were used for evaluation, i.e., training with $5$\%, $10$\%, and $20$\% labeled data, respectively.

In the 2D DCPA model for the ACDC dataset, the U-Net architecture was employed as its backbone. A patch-based training approach was used, where 2D patches with a dimension of $256 \times 256$ were randomly extracted. A batch size of $24$ was used, including $12$ labeled data and $12$ unlabeled samples. The model underwent training for a total of $30,000$ iterations to optimize its performance. It is worth noting that all experimental configurations on the ACDC dataset adhered to the guidelines set by the public benchmark established by \cite{wu2022exploring}, ensuring fair comparisons with other methods.

\subsubsection{Experimental settings} All experiments were conducted on a server equipped with an NVIDIA GeForce RTX $3090$ $24GB$ GPU running Ubuntu $20.04$ and PyTorch $1.12.1$. For the segmentation tasks, we utilized an SGD optimizer with a momentum of $0.9$ and weight decay of $0.0001$. The initial learning rate was set to $0.01$. Regarding the hyperparameter settings, $K$ was defined as $2$, $T$ was set to $0.1$, $H$ was established as $0.5$, and $\alpha$ was assigned a value of $0.75$ for all tasks. In addition, morphological operations were employed to enhance the segmented outcomes, such as selecting the largest connected component as the post-processing module \cite{li2020shape, wu2022exploring}. Four metrics were employed for evaluation: Dice, Jaccard, the $95$\% Hausdorff Distance ($95$HD), and average surface distance (ASD).

\subsection{Experimental results}
To fully assess the effectiveness of our proposed method, we conducted comparisons with both supervised and semi-supervised methods, including V-Net \cite{milletari2016v}, UA-MT \cite{yu2019uncertainty}, SASSNet \cite{li2020shape}, DTC \cite{luo2021semi}, URPC \cite{luo2022semi}, SS-Net \cite{wu2022exploring}, and MC-Net+ \cite{wu2022mutual}. Note that V-Net \cite{milletari2016v} is the supervised learning method while the others are semi-supervised methods. To facilitate fair comparisons, we trained the supervised learning method, i.e., V-Net, with varying a mount of labeled data from $5$\% to $100$\%.

\subsubsection{Results on the Pancreas-CT dataset} Table \ref{tab1} displays quantitative results for the Pancreas-CT dataset, showcasing the superior performance of the proposed method compared to other semi-supervised approaches across all quantitative metrics. With only $5\%$ labeled data, DCPA exhibits a significant performance enhancement compared to the fully-supervised method (V-Net), notably elevating the Dice score from $29.32\%$ to $79.16\%$. Similarly, the Dice score obtained by DCPA with $10\%$ labeled data exceeds that of other excellent semi-supervised methods trained with $20\%$ labeled data. Remarkably, even with only $20$\% labeled data for training, the proposed method achieves quantitative results comparable to the upper bound (V-Net with $100$\% labeled training data). These outcomes vividly highlight the efficacy of the proposed method in harnessing information from unlabeled data to improve segmentation performance.

\begin{table*}[ht]\small
\caption{Comparison with the SOTA methods on the LA dataset.}
\label{tab2}
\begin{center}
\begin{tabular}{llclclccccccc}
\hline
\multicolumn{1}{c}{\multirow{2}{*}{\textbf{Method}}} &  & \multicolumn{3}{c}{\textbf{Scans used}}                                      &  & \multicolumn{7}{c}{\textbf{Metrics}}                                                                \\ \cline{3-5} \cline{7-13} 
\multicolumn{1}{c}{}                                 &  & Labeled                   & \multicolumn{1}{c}{} & Unlabeled                 &  & Dice(\%)$\uparrow$      &           & Jaccard(\%)$\uparrow$    &           & 95HD(voxel)$\downarrow$  &           & ASD(voxel)$\downarrow$   \\ \hline
V-Net                                                &  & 4(5\%)                    & \multicolumn{1}{c}{} & 0                         &  & 52.55          &           & 39.60          &           & 47.05         &           & 9.87          \\
V-Net                                                &  & 8(10\%)                   & \multicolumn{1}{c}{} & 0                         &  & 78.57          &           & 66.96          &           & 21.20         &           & 6.07          \\
V-Net                                                &  & 16(20\%)                  & \multicolumn{1}{c}{} & 0                         &  & 86.96          &           & 77.31          &           & 11.85         &           & 3.22          \\
V-Net                                                &  & 80(All)                   & \multicolumn{1}{c}{} & 0                         &  & 91.71          &           & 84.76          &           & 5.57          &           & 1.59          \\ \hline
UA-MT (MICCAI'19)                                               &  & \multirow{7}{*}{4(5\%)}   &                      & \multirow{7}{*}{76(95\%)} &  & 82.26          &           & 70.98          &           & 13.71         &           & 3.82          \\
SASSNet (MICCAI'19)                                             &  &                           &                      &                           &  & 81.60          &           & 69.63          &           & 16.60         &           & 3.58          \\
DTC (AAAI'21)                                                 &  &                           &                      &                           &  & 81.25          &           & 69.33          &           & 14.90         &           & 3.99          \\
URPC (MIA'22)                                                &  &                           &                      &                           &  & 82.48          &           & 71.35          &           & 14.65         &           & 3.65          \\
SS-Net (MICCAI'22)                                              &  &                           &                      &                           &  & 86.33          &           & 76.15          &           & 9.97          &           & 2.31          \\
MC-Net+ (MIA'22)                                             &  &                           &                      &                           &  & 82.07          &           & 70.38          &           & 20.49         &           & 5.72          \\
\textbf{Ours}                                        &  &                           &                      &                           &  & \textbf{89.34} &           & \textbf{80.79} &           & \textbf{8.13} &           & \textbf{1.69} \\ \hline
UA-MT (MICCAI'19)                                               &  & \multirow{7}{*}{8(10\%)}  &                      & \multirow{7}{*}{72(90\%)} &  & 86.28          & \textbf{} & 76.11          & \textbf{} & 18.71         & \textbf{} & 4.63          \\
SASSNet (MICCAI'19)                                             &  &                           &                      &                           &  & 85.22          & \textbf{} & 75.09          & \textbf{} & 11.18         & \textbf{} & 2.89          \\
DTC (AAAI'21)                                                 &  &                           &                      &                           &  & 87.51          &           & 78.17          &           & 8.23          &           & 2.36          \\
URPC (MIA'22)                                                &  &                           &                      &                           &  & 85.01          &           & 74.36          &           & 15.37         &           & 3.96          \\
SS-Net (MICCAI'22)                                              &  &                           &                      &                           &  & 88.43          &           & 79.43          &           & 7.95          &           & 2.55          \\
MC-Net+ (MIA'22)                                             &  &                           &                      &                           &  & 88.96          &           & 80.25          &           & 7.93          &           & 1.86          \\
\textbf{Ours}                                        &  &                           &                      &                           &  & \textbf{90.91} &           & \textbf{83.39} &           & \textbf{5.53} &           & \textbf{1.65} \\ \hline
UA-MT (MICCAI'19)                                               &  & \multirow{7}{*}{16(20\%)} &                      & \multirow{7}{*}{64(80\%)} &  & 88.74          &           & 79.94          &           & 8.39          &           & 2.32          \\
SASSNet (MICCAI'19)                                             &  &                           &                      &                           &  & 89.16          & \textbf{} & 80.60          & \textbf{} & 8.95          & \textbf{} & 2.26          \\
DTC (AAAI'21)                                                 &  &                           &                      &                           &  & 89.52          &           & 81.22          &           & 7.07          &           & 1.96          \\
URPC (MIA'22)                                                &  &                           &                      &                           &  & 88.74          &           & 79.93          &           & 12.73         &           & 3.66          \\
SS-Net (MICCAI'22)                                              &  &                           &                      &                           &  & 89.86          &           & 81.70          &           & 7.01          &           & 1.87          \\
MC-Net+ (MIA'22)                                             &  &                           &                      &                           &  & 91.07          &           & 83.67          &           & 5.84          &           & 1.99          \\
\textbf{Ours}                                        &  &                           &                      &                           &  & \textbf{91.64} &           & \textbf{84.63} &           & \textbf{5.04} &           & \textbf{1.52} \\ \hline
\end{tabular}
\end{center}
\end{table*}

Fig. \ref{PA} illustrates the visualized segmentation results obtained through training with $5$\% and $10$\% labeled data on the Pancreas-CT dataset, presented in both 2D and 3D perspectives. In the 3D view, the predicted results of other models trained with $5$\% of labeled data exhibit noticeable deformities, while our model's results closely resemble the ground truth. Without a doubt, the results of our model are closer to the ground truth when using $10\%$ of labeled data. But the prediction results of other models primarily suffer from two issues: false positives (e.g., a, d, e) and false negatives (e.g., b, c, f), as indicated by the yellow arrows and green arrows, respectively. The 2D view clearly depicts the disparity between predicted boundaries (green lines) and true boundaries (red shadows). Our model consistently produces smooth and complete segmentation boundaries regardless of $5$\% or $10$\% labeled data were used, highlighting the model's robustness.


\subsubsection{Results on the LA dataset} 
\begin{figure}[!t]
\centerline{\includegraphics[width=\columnwidth]{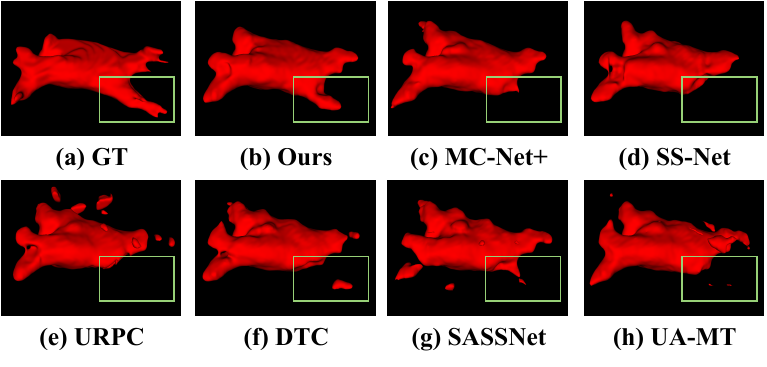}}
\caption{3D views of the segmentation results by different methods on the LA dataset. Note that $5$\% labeled data were used for training.}
\label{LA}
\end{figure}

Our comparisons were extended to encompass an additional 3D dataset, enhancing the assessment of the effectiveness and generalization of our proposed method. Table \ref{tab2} presents the quantitative results. In all semi-supervised settings, the proposed DCPA surpasses other state-of-the-art semi-supervised methods in terms of quantitative metrics. Impressively, even when trained with only $20$\% labeled data, our method produces results nearly comparable to the upper bound established by V-Net, which is trained with a full $100$\% labeled dataset. In addition to the numerical outcomes, the segmentation visualizations in Fig. \ref{LA} underscore the superior performance of our approach. In the highlighted area within Fig. \ref{LA}, it becomes evident that when using only $5$\% of labeled data, the other models exhibit significant instances of missed detections. In contrast, our model offers a more comprehensive representation of the left atrium and preserves finer details.

\begin{figure}[!t]
\centerline{\includegraphics[width=\columnwidth]{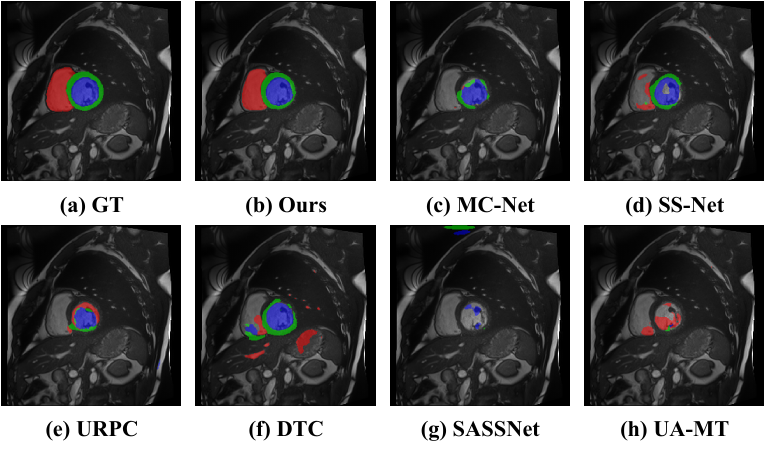}}
\caption{Segmentation results of different method on the 2D ACDC dataset with a $5$\% labeled data. The left ventricle, myocardium, and right ventricle are distinctly color-coded as red, green, and blue, respectively.}
\label{ACDC}
\end{figure}

\begin{table*}[]\small
\caption{Comparison with the SOTA methods on the ACDC dataset.}
\label{tab3}
\begin{center}
\begin{tabular}{llclclccccccc}
\hline
\multicolumn{1}{c}{\multirow{2}{*}{\textbf{Method}}} &  & \multicolumn{3}{c}{\textbf{Scans used}}                                      &  & \multicolumn{7}{c}{\textbf{Metrics}}                                                                \\ \cline{3-5} \cline{7-13} 
\multicolumn{1}{c}{}                                 &  & Labeled                   & \multicolumn{1}{c}{} & Unlabeled                 &  & Dice(\%)$\uparrow$       &           & Jaccard(\%)$\uparrow$    &           & 95HD(voxel)$\downarrow$   &           & ASD(voxel)$\downarrow$    \\ \hline
U-Net                                                 &  & 3(5\%)                    & \multicolumn{1}{c}{} & 0                         &  & 47.83          &           & 37.01          &           & 31.16         &           & 12.62         \\
U-Net                                                 &  & 7(10\%)                   & \multicolumn{1}{c}{} & 0                         &  & 77.34          &           & 66.20          &           & 9.18          &           & 2.45          \\
U-Net                                                 &  & 14(20\%)                  & \multicolumn{1}{c}{} & 0                         &  & 85.15          &           & 75.48          &           & 6.20          &           & 2.12          \\
U-Net                                                 &  & 70(All)                   & \multicolumn{1}{c}{} & 0                         &  & 91.65          &           & 84.93          &           & 1.89          &           & 0.56          \\ \hline
UA-MT (MICCAI'19)                                               &  & \multirow{7}{*}{3(5\%)}   &                      & \multirow{7}{*}{67(95\%)} &  & 50.04          &           & 39.57          &           & 29.31         &           & 9.88          \\
SASSNet (MICCAI'19)                                             &  &                           &                      &                           &  & 57.77          &           & 46.14          &           & 20.05         &           & 6.06          \\
DTC (AAAI'21)                                                 &  &                           &                      &                           &  & 56.90          &           & 45.67          &           & 23.36         &           & 7.39          \\
URPC (MIA'22)                                                &  &                           &                      &                           &  & 55.87          &           & 44.64          &           & 13.60         &           & 3.74          \\
SS-Net (MICCAI'22)                                              &  &                           &                      &                           &  & 65.82          &           & 55.38          &           & 6.67          &           & 2.28          \\
MC-Net+ (MIA'22)                                             &  &                           &                      &                           &  & 64.77          &           & 54.06          &           & 12.39         &           & 3.45          \\
\textbf{Ours}                                        &  &                           &                      &                           &  & \textbf{85.86} &           & \textbf{76.13} &           & \textbf{5.83} &           & \textbf{2.03} \\ \hline
UA-MT (MICCAI'19)                                               &  & \multirow{7}{*}{7(10\%)}  &                      & \multirow{7}{*}{63(90\%)} &  & 81.58          & \textbf{} & 70.48          & \textbf{} & 12.35         & \textbf{} & 3.62          \\
SASSNet (MICCAI'19)                                             &  &                           &                      &                           &  & 84.14          & \textbf{} & 74.09          & \textbf{} & 5.03          & \textbf{} & 1.40          \\
DTC (AAAI'21)                                                 &  &                           &                      &                           &  & 82.71          &           & 72.14          &           & 11.35         &           & 2.99          \\
URPC (MIA'22)                                                &  &                           &                      &                           &  & 81.77          &           & 70.85          &           & 5.04          &           & 1.41          \\
SS-Net (MICCAI'22)                                              &  &                           &                      &                           &  & 86.78          &           & 77.67          &           & 6.07          &           & 1.40          \\
MC-Net+ (MIA'22)                                             &  &                           &                      &                           &  & 87.10          &           & 78.06          &           & 6.68          &           & 2.00          \\
\textbf{Ours}                                        &  &                           &                      &                           &  & \textbf{88.10} &           & \textbf{79.45} &           & \textbf{1.86} &           & \textbf{0.57} \\ \hline
UA-MT (MICCAI'19)                                               &  & \multirow{7}{*}{14(20\%)} &                      & \multirow{7}{*}{56(80\%)} &  & 85.87          &           & 76.78          &           & 5.06          &           & 1.54          \\
SASSNet (MICCAI'19)                                             &  &                           &                      &                           &  & 87.04          & \textbf{} & 78.13          & \textbf{} & 7.84          & \textbf{} & 2.15          \\
DTC (AAAI'21)                                                 &  &                           &                      &                           &  & 86.28          &           & 77.03          &           & 6.14          &           & 2.11          \\
URPC (MIA'22)                                                &  &                           &                      &                           &  & 85.07          &           & 75.61          &           & 6.26          &           & 1.77          \\
SS-Net (MICCAI'22)                                              &  &                           &                      &                           &  & 88.10          &           & 79.70          &           & 3.34          &           & 0.95          \\
MC-Net+ (MIA'22)                                             &  &                           &                      &                           &  & 88.51          &           & 80.19          &           & 5.35          &           & 1.54          \\
\textbf{Ours}                                        &  &                           &                      &                           &  & \textbf{89.19} &           & \textbf{81.08} &           & \textbf{1.71} &           & \textbf{0.46} \\ \hline
\end{tabular}
\end{center}
\end{table*}

\subsubsection{Results on the ACDC dataset} 
The proposed method was further extended to address 2D multi-class segmentation tasks. Quantitative metrics were averaged across three classes: the myocardium, left ventricle, and right ventricle, and the results are presented in Table \ref{tab3}. The results demonstrate that in various semi-supervised settings, the proposed method consistently ranks at the top across Dice, Jaccard, $95$HD, and ASD metrics, outperforming competing semi-supervised methods. Regardless of the percentage of labeled data used, the proposed method consistently achieves a Dice score above $85.00$\%, affirming its stability and efficacy. Moreover, the visualized results also clearly demonstrate the superior performance of the proposed method over the competing methods. As shown in Fig. \ref{ACDC}, competing methods encounter challenges in accurately segmenting the left ventricle, and defining precise boundaries for the myocardium and right ventricle remains a challenge for several of them.

Overall, experimental results across three segmentation datasets, spanning both 2D and 3D imaging, highlight the proposed method's superiority over current state-of-the-art competing methods. Notably, the proposed method, with just $20$\% labeled data, can yield results on par with the supervised learning method using a full $100$\% labeled dataset. These results indicate the proposed method's capability in harnessing useful information from unlabeled data.

\begin{table*}[!t]\small
\begin{center}
\begin{threeparttable}
\caption{Ablation results the Pancreas-CT dataset to evaluate the contribution of each component of the proposed method.}
\begin{tabular}{ccccccccccccccc}
\hline
\multicolumn{3}{c}{\textbf{Scans used}}                  & \multicolumn{4}{c}{\textbf{Method}} &  & \multicolumn{7}{c}{\textbf{Metrics}}                                                                \\ \cline{1-7} \cline{9-15} 
Labeled                   &  & Unlabeled                 & DD     & PL     & MT     & Mixup    &  & Dice(\%)       &           & Jaccard(\%)    &           & 95HD(voxel)   &           & ASD(voxel)    \\ \hline
\multirow{6}{*}{3(5\%)}   &  & \multirow{6}{*}{59(95\%)} &        &        &        &          &  & 29.32          &           & 19.61          &           & 43.67         &           & 15.42         \\
                          &  &                           & \checkmark      &        &        &          &  & 36.77          &           & 24.47          &           & 46.34         &           & 16.41         \\
                          &  &                           & \checkmark      & \checkmark      &        &          &  & 44.55          &           & 30.55          &           & 55.97         &           & 21.20         \\
                          &  &                           & \checkmark      & \checkmark      & \checkmark      &          &  & 62.65          &           & 47.59          &           & 34.57         &           & 12.08         \\
                          &  &                           & \checkmark      & \checkmark      &        & \checkmark        &  & 69.07          &           & 53.64          &           & 26.08         &           & 7.55          \\
                          &  &                           & \checkmark      & \checkmark      & \checkmark      & \checkmark        &  & \textbf{79.16} &           & \textbf{65.86} &           & \textbf{5.88} &           & \textbf{1.69} \\ \hline
\multirow{6}{*}{6(10\%)}  &  & \multirow{6}{*}{56(90\%)} &        &        &        &          &  & 54.94          & \textbf{} & 40.87          & \textbf{} & 47.48         & \textbf{} & 17.43         \\
                          &  &                           & \checkmark      &        &        &          &  & 66.56          & \textbf{} & 52.10          & \textbf{} & 16.80         & \textbf{} & 2.32          \\
                          &  &                           & \checkmark      & \checkmark      &        &          &  & 71.12          &           & 56.41          &           & 37.71         &           & 11.52         \\
                          &  &                           & \checkmark      & \checkmark      & \checkmark      &          &  & 79.38          &           & 66.47          &           & 6.83          &           & 1.73          \\
                          &  &                           & \checkmark      & \checkmark      &        & \checkmark        &  & 78.02          &           & 64.44          &           & 9.61          &           & 3.02          \\
                          &  &                           & \checkmark      & \checkmark      & \checkmark      & \checkmark        &  & \textbf{81.34} &           & \textbf{68.97} &           & \textbf{5.40} &           & \textbf{1.69} \\ \hline
\multirow{6}{*}{12(20\%)} &  & \multirow{6}{*}{50(80\%)} &        &        &        &          &  & 71.52          &           & 57.68          &           & 18.12         &           & 5.41          \\
                          &  &                           & \checkmark      &        &        &          &  & 76.74          & \textbf{} & 63.64          & \textbf{} & 9.52          & \textbf{} & 2.58          \\
                          &  &                           & \checkmark      & \checkmark      &        &          &  & 79.17          &           & 66.23          &           & 9.79          &           & 2.56          \\
                          &  &                           & \checkmark      & \checkmark      & \checkmark      &          &  & 81.38          &           & 69.04          &           & 8.71          &           & 3.71          \\
                          &  &                           & \checkmark      & \checkmark      &        & \checkmark        &  & 82.64          &           & 70.75          &           & 5.31          &           & 1.34          \\
                          &  &                           & \checkmark      & \checkmark      & \checkmark      & \checkmark        &  & \textbf{82.89} &           & \textbf{71.17} &           & \textbf{4.54} &           & \textbf{1.23} \\ \hline
\end{tabular}
\label{tab4}
\begin{tablenotes}
\item "DD" denotes the dual-decoder consistency, "PL" the pseudo-labels, and "MT" the mean-teacher model. A check mark denotes introducing the associated component in the proposed method.
\end{tablenotes}
\end{threeparttable}
\end{center}
\end{table*}

\section{Discussion}
\subsection{Ablation Studies}
In this section, ablation studies were performed on 1) DCPA components, 2) the number of decoders $N$, 3) the amount of weak data augmentation on unlabeled data $K$, and 4) temperature $T$.

\subsubsection{Proposed model components} 
Experiments were conducted on the Pancreas-CT dataset to evaluate the contributions of the various components of the proposed method. The quantitative results under different settings ($5$\%, $10$\%, and $20$\% labeled data) are listed in Table \ref{tab4}. From the experimental results, we empirically highlight that in semi-supervised learning, each individual component distinctly contributes to performance improvements. Furthermore, the peak of performance is realized when these elements work in synergy.

\begin{figure}[!t]
\centerline{\includegraphics[width=\columnwidth]{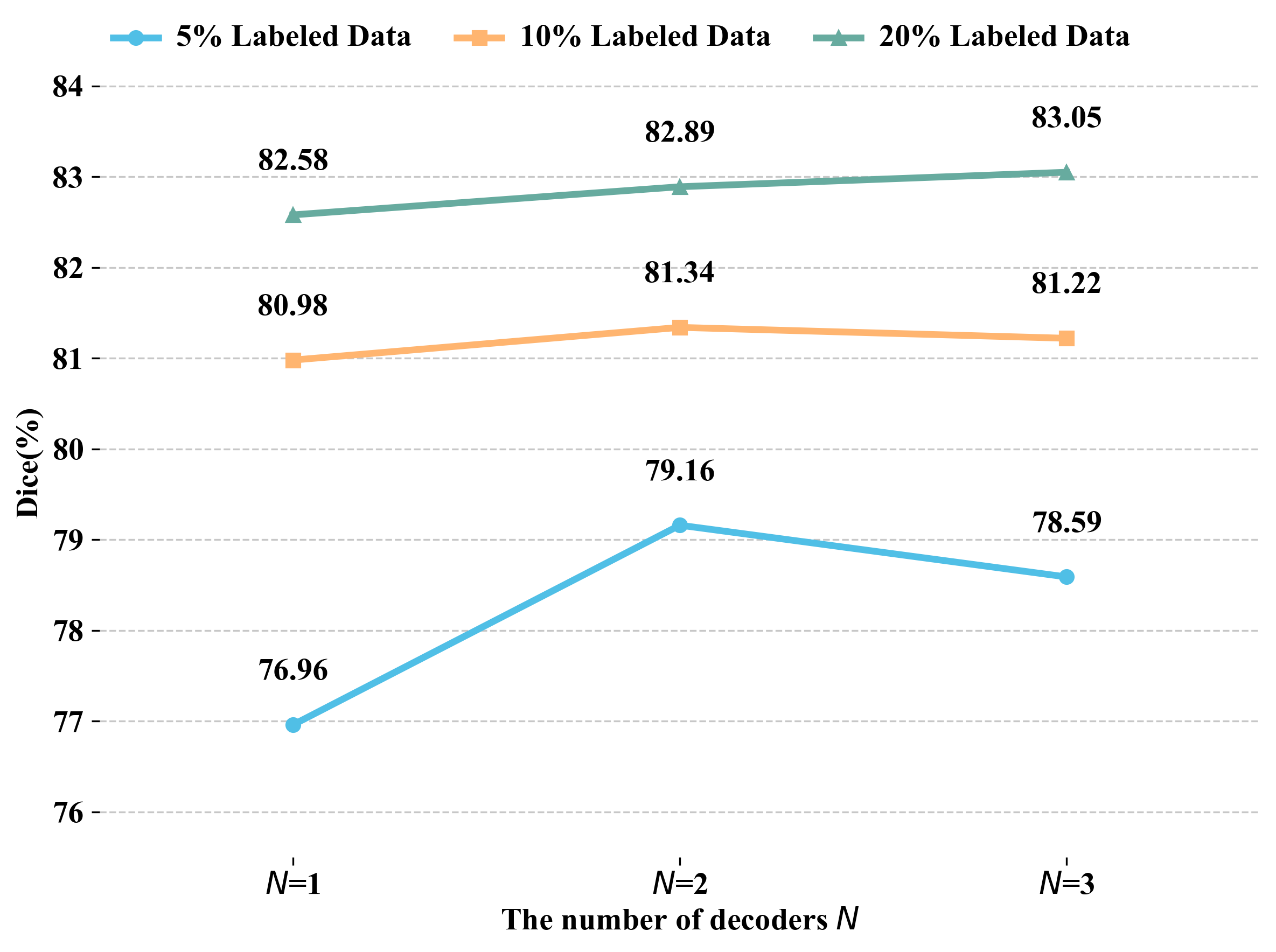}}
\caption{Dice score variation versus the number of decoders on the Pancreas-CT dataset.}
\label{decoder}
\end{figure}

\subsubsection{Number of decoders}
The impact of the number of decoders ($N$) on the Pancreas-CT dataset was also examined across three semi-supervised settings. In our experiments, for $N=1$, we employed the transposed convolutional layer for upsampling. For $N=2$, a second decoder was introduced, using linear interpolation for upsampling. Similarly, in the vein of MC-Net+, $N=3$ added a third decoder with nearest interpolation for upsampling. It's worth noting that consistency losses between different decoders were incorporated for both $N=2$ and $N=3$. The results, illustrated in Fig. \ref{decoder}, show that combining consistency loss with multiple decoders can boost performance. While $N=2$ achieves a peak performance of $83.05$\% with $20$\% labeled data, it falls in scenarios with fewer annotations and also incurs higher training costs. 
For optimal performance, setting $N=2$ in the proposed method is recommended.

\begin{figure}[!t]
\centerline{\includegraphics[width=\columnwidth]{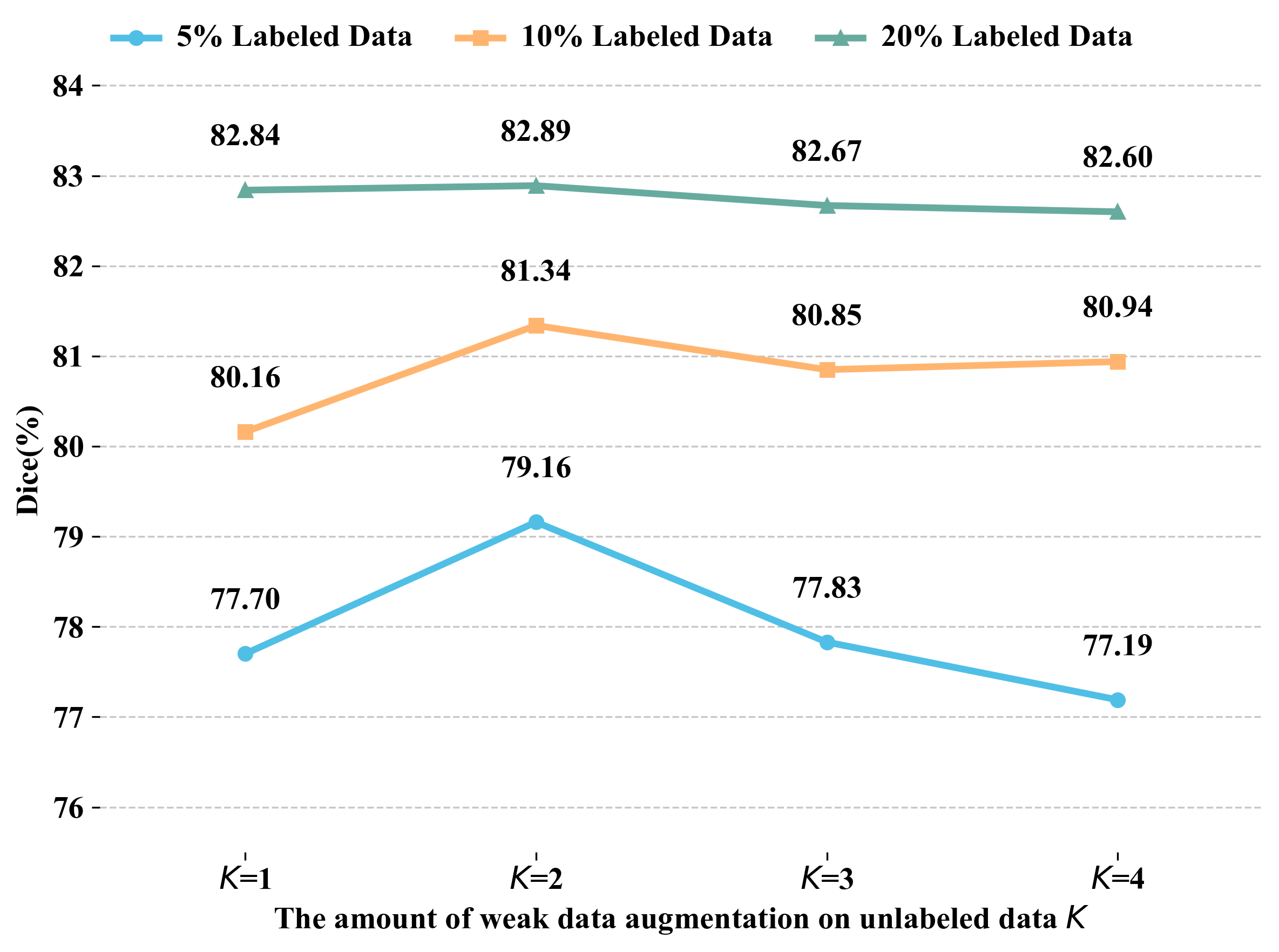}}
\caption{Dice score versus the number of weak data augmentation on the Pancreas-CT dataset.}
\label{K}
\end{figure}

\subsubsection{Number of weak data augmentation} 
Evaluations were performed on the Pancreas-CT dataset in three semi-supervised scenarios to analyze the influence of the number of weak data augmentations ($K$) applied to unlabeled data. For $K=1$, the predictions of the teacher model are directly used as pseudo-labels for the unlabeled data. Otherwise, the average of the predictions of the teacher model serves as the pseudo-labels. As depicted in Fig. \ref{K}, under the $5$\% labeled data scenario, the performance of our proposed method exhibits heightened sensitivity to variations in $K$. However, as the amount of labeled data increases, the effect of altering $K$ diminishes. Particularly, a consistent peak performance is observed with $K=2$ across all settings. Hence, selecting $K=2$ for network training was our choice.

\subsubsection{Sharpening temperatures}
The inclusion of a sharpening function in the proposed method is intended to improve the clarity of segmentation boundaries and reduce the presence of blurry features. Evaluations were conducted on three datasets with only 10\% labeled data to evaluate the impact of the temperature coefficient, denoted as $T$, in the sharpening function. The results shown in Fig. \ref{T} indicate the robustness of the proposed method to the temperature coefficient. Note that a relatively high $T$ value makes the proposed method prone to blurry features, diminishing its effectiveness. Conversely, a relatively low $T$ value expose yields the pitfalls of incorrect predictions. The proposed method achieves its peak performance at $T=0.1$. Based on these observations, $T=0.1$ was chosen as the temperature coefficient for the sharpening function.

\subsection{Limitations and Future Works} 
While the proposed method allows promising results in medical image segmentation, its dual-decoder consistency has been primarily centered on exploring inter-decoder differences, with a somewhat narrow exploration of upsampling strategies. Moving forward, several interesting research directions emerge:

\begin{figure}[!t]
\centerline{\includegraphics[width=\columnwidth]{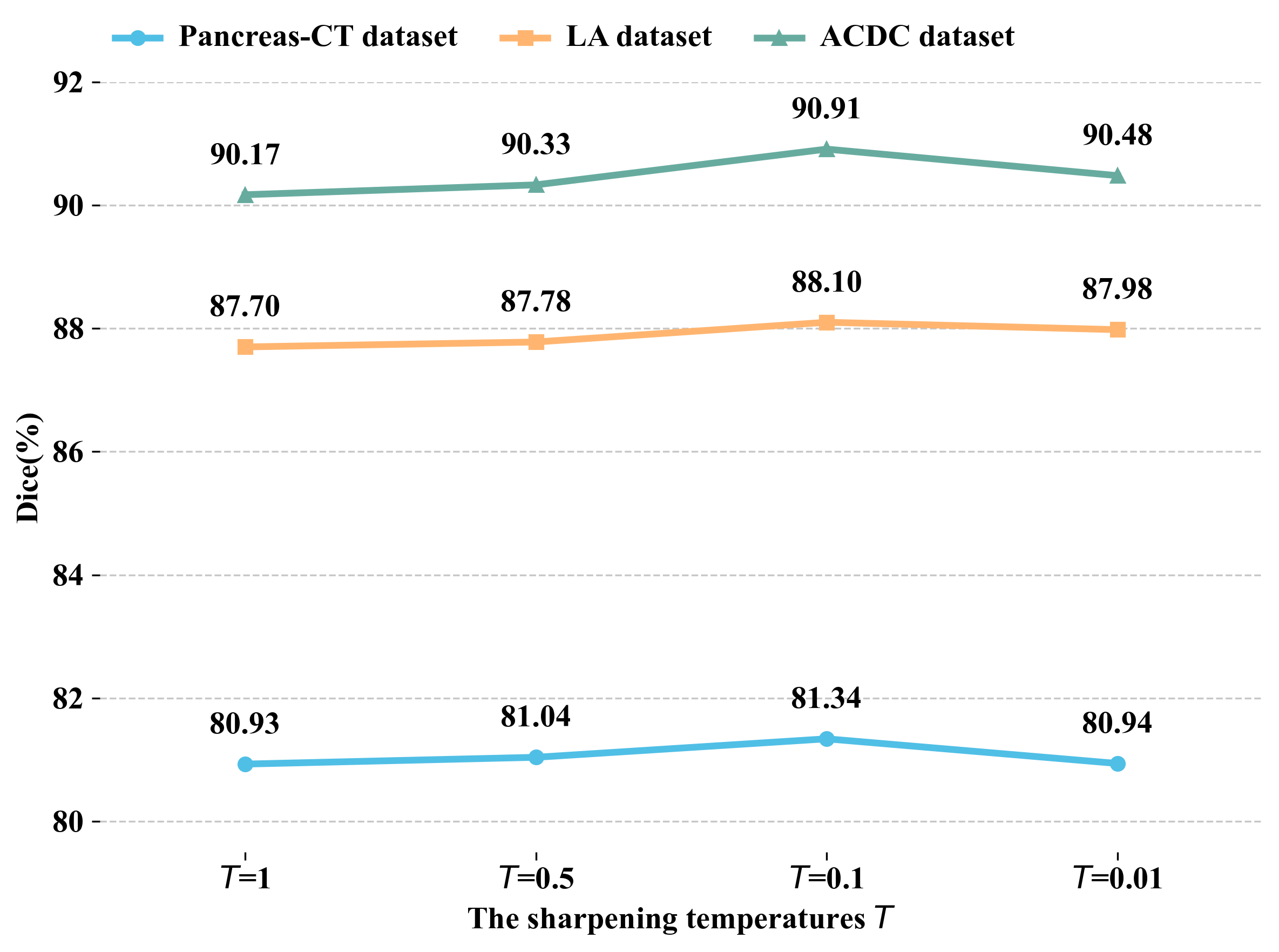}}
\caption{Dice scores versus sharpening temperature on three datasets.}
\label{T}
\end{figure}

\begin{itemize}
	\item Model Disparities: Investigating the differences between various models, such as Transformers and CNNs, might allow us to harness the unique strengths of each architecture.

	\item Synergistic Approaches: Combining our Pseudo-Label Guided Data Augmentation method with other advanced techniques, like Generative Adversarial Networks \citep{goodfellow2014generative} and Meta-Learning \citep{ren2018meta}, can potentially enhance the performance and resilience.

	\item Expanded Applicability: Broadening the scope of our method to encompass wider medical image domains, such as semi-supervised medical image classification, could capitalize on the power of limited labeled data, especially in distinguishing between benign and malignant cases.
\end{itemize}

\section{Conclusion}
To tackle the challenge of medical image segmentation task under limited labeled data, we introduced the DCPA model, a semi-supervised approach built upon the mean-teacher framework. This model employs a dual-decoder architecture, utilizing the teacher model's predictions as pseudo-labels for unlabeled data. To enhance the training process, we integrated a data augmentation technique, amalgamating labeled and unlabeled datasets. Additionally, constraints were applied during training to ensure consistent predictions between the dual-decoders.

Empirical experiments demonstrate the superiority of the proposed method, outperforming six state-of-the-art models across diverse scenarios on three public datasets. Its robust performance, particularly with just $5$\% labeled data, highlights its potential in data-scarce medical image segmentation. This research not only establishes the effectiveness of the proposed method but also suggests a promising direction for future semi-supervised segmentation endeavors.

\section*{Acknowledgments}
This work was supported by National Natural Science Foundation of China under Grant 62171133, in part by the Artificial Intelligence and Economy Integration Platform of Fujian Province, the Fujian Health Commission under Grant 2022ZD01003, and the Guiding Projects of Fujian Provincial Technology Research and Development Program under Grant 2022Y0023.



 \bibliographystyle{elsarticle-harv} 
 \bibliography{ref}





\end{document}